\documentclass[a4paper,11pt]{article}
\pdfoutput=1 

\usepackage{jinstpub} 
\usepackage{graphicx}
\usepackage{siunitx}
\usepackage{amsmath}

\title{Modelling the radiation pattern of a dual circular polarization system}

\author[1]{S. Realini\note{Corresponding author.},}
\author{C. Franceschet,}
\author{A. Mennella}
\affiliation{Universit\`a degli Studi di Milano, via Celoria 16, Milano, Italy}

\emailAdd{sabrina.realini@unimi.it}

\abstract{We present the electromagnetic model of a dual circular polarization antenna-feed system, consisting of a corrugated feedhorn, a polarizer and an orthomode transducer. This model was developed for the passive front-end implemented in the Q-band receivers of the STRIP instrument of the Large Scale Polarization Explorer experiment. Its applicability, however, is completely general. The model has been implemented by superposing the response of two linearly polarized feedhorns with a $\pi$/2 phase difference, thus taking into account for the effect of the polarizer, that behaves differently for the two polarizations of the incoming electric field. The model has been verified by means of radiation pattern measurements performed in the anechoic chamber at the Physics Department of the University of Milan. We measured both arms of the orthomode transducer, in order to check that the diagram at one port is the 90$^{\circ}$ rotation of the diagram at the other port. Simulations and measurements show an agreement at the level of fraction of a dB up to the first sidelobes, thus confirming the model.}


\begin{document}

\maketitle
\flushbottom

\section{Introduction}
\label{intro}
Discrimination of polarization to ever more stringent levels is a fundamental characteristic of many instruments for cosmological and astrophysical observations in the radio to sub-mm range. To achieve this goal, several instruments implement a dual circular polarization (DCP) systems, together with correlation radiometric architectures capable of extracting the polarization information. 
One example is the STRIP instrument of the Large Scale Polarization Explorer\cite{STRIP} that will observe the microwave sky with an array of forty-nine receivers in the Q-band. Each receiver implements a DCP chain by means of a corrugated feedhorn followed by a polarizer and an orthomode transducer (OMT). Similarly, the future Q-band array of the Sardinia Radio Telescope\cite{SRT} will be populated by nineteen radiometric chains with DCP assemblies in their passive front-end. The selected architecture provides the best trade-off between operating bandwidth (20\%) and electrical performance. 

With DCP systems, however, one cannot describe the radiation properties with the conventional concepts of co- and cross-polar radiation patterns, because the co-polar response is split between the left (LHCP) and right (RHCP) circularly polarized components, and the cross-polar response depends on the ratio of LHCP over RHCP.

This has an impact on the characterization of the spurious polarization introduced by the optics based on DCP systems, which requires precise measurements supported by reliable simulations with a realistic model of the beam.

These considerations motivate our work in which we set up a realistic analytical electromagnetic model for a dual polarization chain, consisting of a feedhorn, a polarizer and an orthomode transducer, and compare the results with measurements.

\section{The dual circular-polarization waveguide system}
\label{sec:dc_pol_sys}
Our system consists of a corrugated feedhorn, a grooved polarizer and a turnstile-junction orthomode transducer\cite{polarizer,omt}.
The circular polarization assembly is fed by a linearly polarized signal that can be modelled as a plane wave. The system splits the incoming radiation into two components proportional to $\frac{E_x + i E_y}{\sqrt{2}}$ and $\frac{E_x - i E_y}{\sqrt{2}}$, where $E_x$ and $E_y$ represent the two linear polarizations of the signal, as decomposed in the polarizer basis. In the notation of the field components we are implicitly assuming the time dependence $e^{-i\omega t}$.

\begin{figure}
  \centering
  \includegraphics[width=0.6\textwidth]{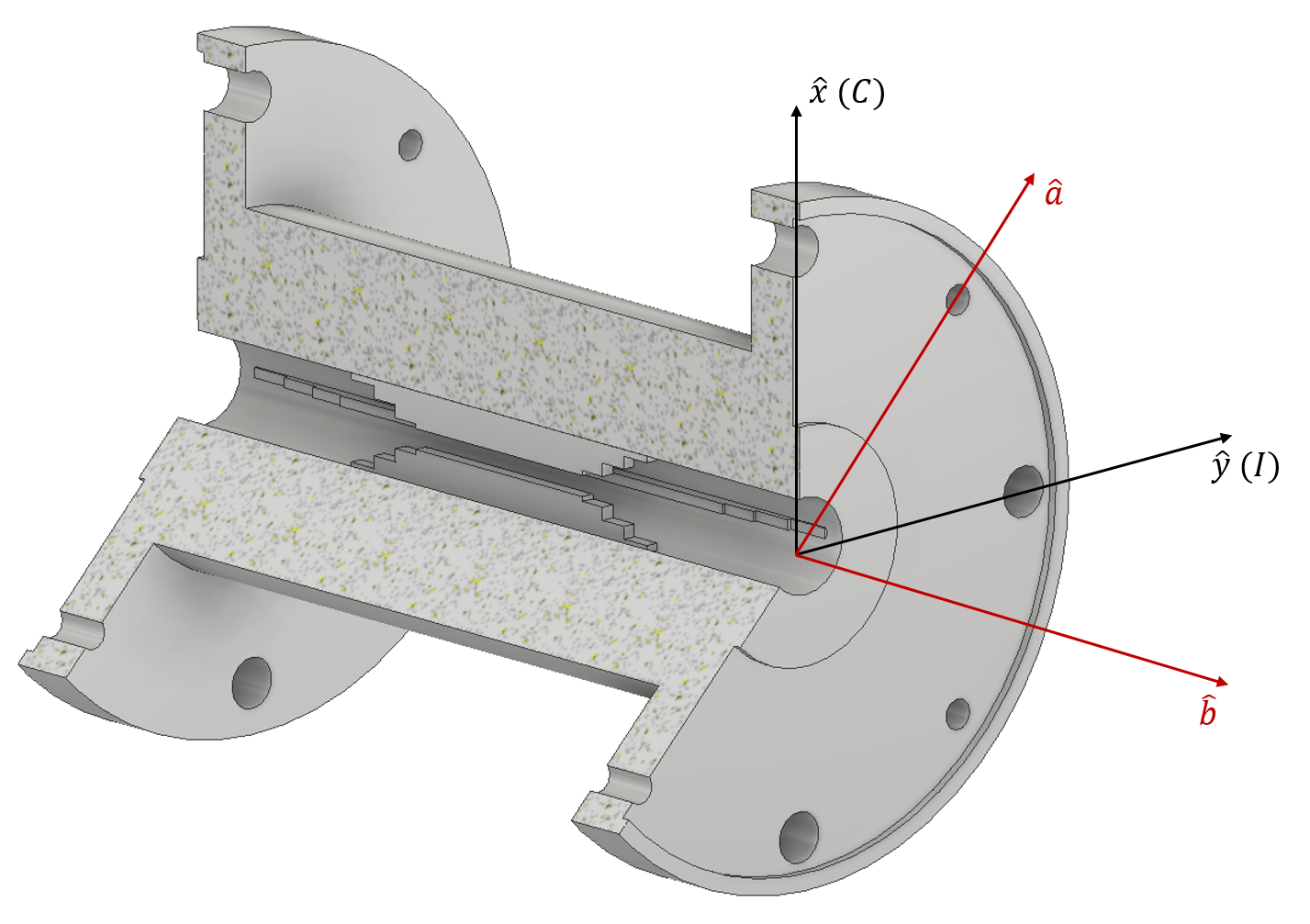}
\caption{Scheme of the grooved polarizer with its inductive ($y$) and capacitive ($x$) axis. The mutual position of the OMT basis is reported in red.}
\label{fig:omt_basis}       
\end{figure}

A key role is played by the polarizer, which converts a linearly polarized field into a circular polarized one, which is subsequently split by the orthomode transducer (OMT).
The polarizer is realized through the insertion of grooves along the waveguide. These allow us to fix a system of coordinates that we use to define Cartesian component of the vector $\vec{E}$(Fig.~\ref{fig:omt_basis}). 

Let us consider a linearly polarized electric field propagating through the polarizer. The grooves behave differently for the two electric field components: the component along the $\hat{y}$ axis, called \emph{inductive axis}, is shifted in phase by $\pi/2$, while the one along the $\hat{x}$ axis, called \emph{capacitive axis}, propagates without any phase variation. 

Following the polarizer we find an OMT oriented at $45^{\circ}$ with respect to the polarizer basis, so that any signal $\vec{E} =(E_x, E_y)$ is decomposed according to
\begin{equation}
\hat{E_a}=\frac{1}{\sqrt{2}}(\hat{E_x}+\hat{E_y}),\qquad\quad
\hat{E_b}=\frac{1}{\sqrt{2}}(\hat{E_x}-\hat{E_y})
\label{omt_rel}
\end{equation}
If we also consider the effect of the polarizer then it is apparent that the fields in the two OMT arms are the left and right circular polarization components defined by
\begin{equation}
\hat{E_l}=\frac{1}{\sqrt{2}}(\hat{E_x}+i\hat{E_y}),\qquad \quad
\hat{E_r}=\frac{1}{\sqrt{2}}(\hat{E_x}-i\hat{E_y})
\end{equation}

\subsection{Scattering matrices}
\label{sec:scat_matrix}
In this section we describe the DCP system using the scattering matrix formalism.
The polarizer has two physical ports, but four electrical ports are required in order to take into account the two polarizations of the mode. As a consequence, the scattering matrix is a $ 4\times4$ matrix, that can be divided into four blocks related to the scattering parameters. Generally, the scattering matrix block that relates the initial state and the final state of a physical system is called $S_{21}$ and is a $ 2\times2$.
In the ideal case, where insertion losses can be neglected and there is no forward cross-coupling between modes, $S_{21}$ is diagonal and contains only phase shift terms along the base axis. In the following mathematical modelling of the dual polarization chain we will consider ideal devices.

If $\vec{E}=E_x\hat{x}+ E_y\hat{y}$ is the incident field, then the scattered field at the second physical port of the polarizer is given by
\begin{equation}
\begin{pmatrix}
E^{\text{out}}_{x} \\
E_y^{\text{out}}
\end{pmatrix}
=
\begin{pmatrix}
e^{i\phi_{x}} & 0 \\
0 & e^{i\phi_{y}}
\end{pmatrix}
\begin{pmatrix}
E_x \\
E_y
\end{pmatrix},
\end{equation}
and, if the phase shifts are $\phi_x=0$ and $\phi_y$=\SI{90}{\degree}, the polarizer scattering matrix becomes
\begin{equation}
S^{\text{POL}}= 
\begin{pmatrix}
1 & 0 \\
0 & i
\end{pmatrix}.
\end{equation}
The OMT, connected to the polarizer, splits the incoming radiation and its base is rotated by \SI{45}{\degree} w.r.t. the polarizer principal axis. According to equation \ref{omt_rel}, the OMT transfer matrix can be written as
\begin{equation}
S^{\text{OMT}}=\frac{1}{\sqrt{2}} 
\begin{pmatrix}
1 & 1 \\
1 & -1
\end{pmatrix}.
\end{equation}

Once we know $S^{\text{POL}}$ and $S^{\text{OMT}}$, we can write the whole chain scattering matrix as a simple matrix product:
\begin{equation}
S^{\text{TOT}}=S^{\text{OMT}}\cdot S^{\text{POL}}=\frac{1}{\sqrt{2}} 
\begin{pmatrix}
1 & i \\
1 & -i
\end{pmatrix}.
\end{equation}
If the incoming vector is $\begin{pmatrix}
E_x \\
E_y
\end{pmatrix}$, then, using the whole system scattering matrix, we find that the outputs at the OMT ports are $\frac{E_x+iE_y}{\sqrt{2}}$ and $\frac{E_x-iE_y}{\sqrt{2}}$. 

Given the scattering matrix, we can also see what happens to incoming radiation polarized in different directions, in terms of unit vectors:
\begin{itemize}
\item linear polarization along $\hat{x}$ 
$
\quad
\begin{pmatrix}
1 \\
0
\end{pmatrix}
\xrightarrow{pol}
\begin{pmatrix}
1 \\
0
\end{pmatrix}
\xrightarrow{omt}
\frac{1}{\sqrt{2}}
\begin{pmatrix}
1 \\
1
\end{pmatrix}
$
\item linear polarization along $\hat{y}$ 
$
\quad
\begin{pmatrix}
0 \\
1
\end{pmatrix}
\xrightarrow{pol}
\begin{pmatrix}
0 \\
i
\end{pmatrix}
\xrightarrow{omt}
\frac{1}{\sqrt{2}}
\begin{pmatrix}
i \\
-i
\end{pmatrix}
$

\item linear polarization at $\theta$ 
$
\quad
\begin{pmatrix}
\sin\theta \\
\cos\theta
\end{pmatrix}
\xrightarrow{pol}
\begin{pmatrix}
\sin\theta \\
i\cos\theta
\end{pmatrix}
\xrightarrow{omt}
\frac{1}{\sqrt{2}}
\begin{pmatrix}
\sin\theta+i\cos\theta \\
\sin\theta-i\cos\theta
\end{pmatrix}
$
\item left-hand circular polarization
$
\quad
\frac{1}{\sqrt{2}}
\begin{pmatrix}
1 \\
i
\end{pmatrix}
\xrightarrow{pol}
\frac{1}{\sqrt{2}}
\begin{pmatrix}
1 \\
-1
\end{pmatrix}
\xrightarrow{omt}
\begin{pmatrix}
0 \\
1
\end{pmatrix}
$
\item right-hand circular polarization
$
\quad
\frac{1}{\sqrt{2}}
\begin{pmatrix}
1 \\
-i
\end{pmatrix}
\xrightarrow{pol}
\frac{1}{\sqrt{2}}
\begin{pmatrix}
1 \\
1
\end{pmatrix}
\xrightarrow{omt}
\begin{pmatrix}
1 \\
0
\end{pmatrix}
$
\end{itemize}

The outgoing vector components represent the expected output at the OMT ports.
According to examples above, in the case of a linearly polarized input, the output has the same amplitude at the ports, but the phase difference changes. Left and right circular polarization cases are reported just for completeness and they produce linear polarization.

Considering the transmitting system, if we excite one port only of the OMT, the radiation emitted by the horn will be circularly polarized. 
If we excite both ports, the output will be a combination of left and right circular polarizations. We recall that we can always describe a linearly polarized radiation as the combination between a left and a right circular polarized radiation.

\section{Modelling the dual circular-polarization system}
\label{sec:model}
Here we define a way to model the electromagnetic response of the dual circular polarization chain described in section \ref{sec:dc_pol_sys}, using the software GRASP by TICRA\footnote{www.ticra.com/software/grasp/}. We start considering a simpler system: a feed-horn connected to a single-mode rectangular waveguide, with the electric field parallel to the short side in the fundamental propagation mode. As a consequence, we can define the system orientation in terms of E- and H-plane\footnote{In a linearly polarized antenna the E- and H-planes are the planes that contain, respectively, the E and H vectors at the antenna aperture.}.

In a dual circular polarization system, instead, we cannot use the same description adopted for linearly polarized horns, as we do not have a rectangular waveguide to select E- and H-plane in a trivial way. In this case, the principal base of the system is the polarizer, with the principal axes along the inductive and capacitive directions. If we set the $x$ axis along one of these two directions we have that the electric field can have two possible orientations: one parallel to $x$, that we denote with $\phi=0$ and one perpendicular to $x$, indicated with $\phi=\pi/2$. The angle $\phi$ is the angle between $x$ and the E vector.

\begin{figure}
\centering
\includegraphics[width=0.7\textwidth]{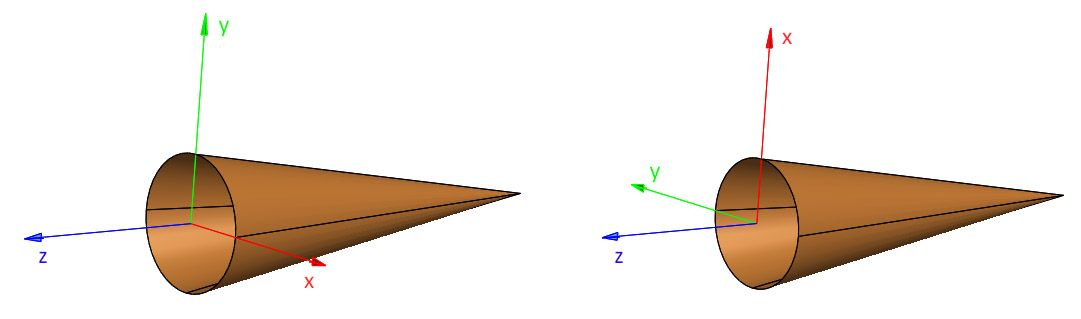}
\caption{The figure shows two feedhorns with reference systems rotated by $\pi/2$ one with respect to the other. To model the dual circular polarization system in GRASP we overlap these two horns in space and introduce a relative phase difference of $\pi/2$.}
\label{fig:feed_cop}
\end{figure}

Because both fields are allowed to propagate through the system, we can model the dual circular polarization horn by overlapping two linearly polarized feedhorns, with their E-planes aligned with the polarizer axis, and with a $\pi/2$ phase difference (see Fig.~\ref{fig:feed_cop}).

\section{Simulations with a corrugated feedhorn} 

For the analysis of the radiation pattern of the dual circular polarization waveguide system, we considered a corrugated feedhorn. The radiation pattern of the feedhorn has been computed using the software SRSR-D\footnote{Software for a Rigorous Simulation of Radiating Structures with Symmetry of Revolution – Dielectric}, which provides reliable and accurate simulations. SRSR-D provides a rigorous simulation of the electromagnetic performances of any structure with symmetry of revolution consisting of conducting parts and homogeneous dielectric domains\cite{srsr}. Simulations take into account the effect of return loss, which is lower than 30 dB, but we do not consider possible ohmic losses. This is due to the fact that we consider a perfect electric conductor.

\begin{figure}
\centering
\includegraphics[width=0.7\textwidth]{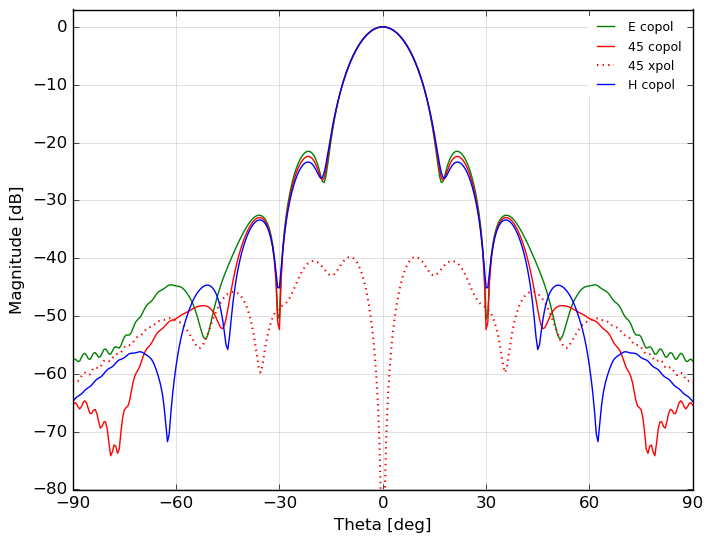}
\caption{Simulated radiation pattern of the corrugated feedhorn at \SI{41.5}{\giga\hertz}.}
\label{fig:cut_srt}
\end{figure}

In order to perform a simulation of the dual circular polarization system, we defined in GRASP two feedhorns and their respective coordinate systems, using the tabulated pattern shown in Fig.~\ref{fig:cut_srt} for both feedhorns. We rotated one feedhorn coordinate system by \SI{90}{\degree} with respect to the other and we introduced a phase difference of \SI{90}{\degree}. In this way, we can reproduce the effect of the polarizer, which introduces a phase-shift on the incoming radiation along one axis. 
For practical reasons, we call \textit{base feed} the feedhorn aligned with the coordinate system used for the analysis and \textit{rotated feed} the feedhorn with the rotated coordinate system and phase-shifted.

The main beams are computed in the far-field region, where the field has no radial component and it can be decomposed along any two orthogonal polarization vectors, which are orthogonal to the far field direction\cite{Ticra:2008}. In our simulations, the definition of the polarization vector is given according to Ludwig's 3rd definition\cite{Ludwig}. The polarizations are parallel to $x$ and $y$, respectively, so that we can call co-polar the beam relative to the orientation of the $x$-axis and cross-polar the one along the $y$-axis.
According to this definition, the main beam computed in the base feed coordinate system show that a $90^{\circ}$ rotation of the feedhorn along its axis causes an exchange of its co-polar and cross-polar components. 

\begin{figure}
\centering
\includegraphics[width=.48\textwidth]{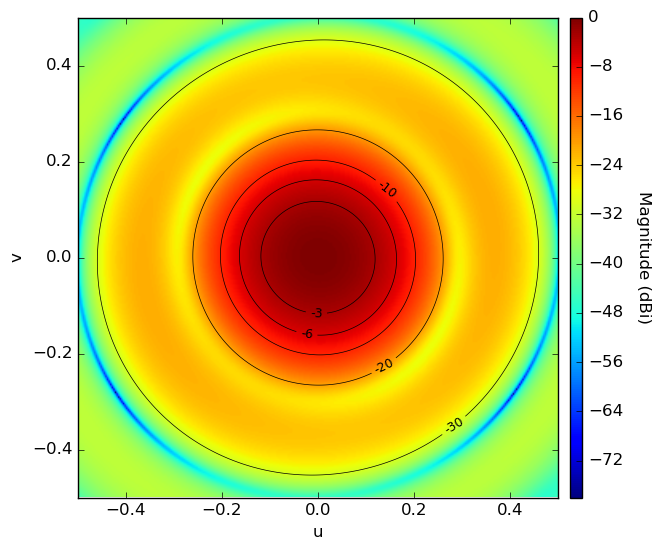}
\includegraphics[width=.48\textwidth]{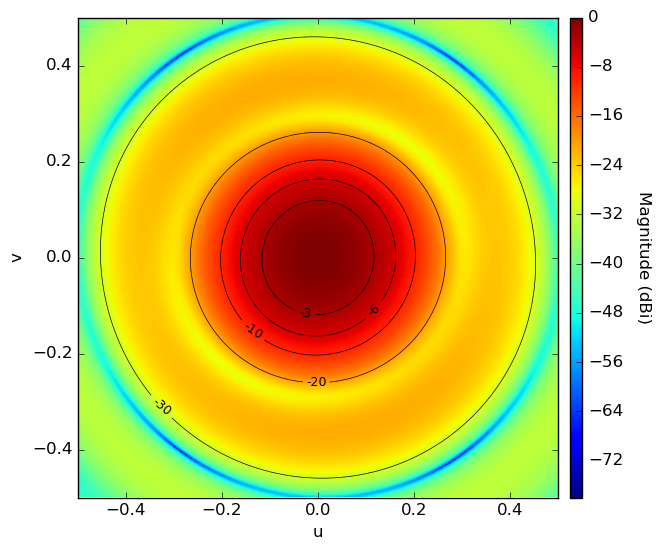}
\caption{Co-polar (\textit{left}) and cross-polar (\textit{right}) peak-normalized components of the double-feed system at 41.5 GHz. Co-polar and cross-polar peak values are the same.}
\label{fig:df_grd}
\end{figure}

We have simulated the radiation pattern using both feedhorns as source and Fig.~\ref{fig:df_grd} reports the co-polar and cross-polar components represented into \textit{uv}-grids for the so-called \textit{double-feed system}.
Considering double-feed system cuts, we find that the co-polar cut at $\phi=0^{\circ}$ coincides with the cross-polar cut at $\phi=90^{\circ}$ and the co-polar cut at $\phi=90^{\circ}$ coincides with the cross-polar cut at $\phi=0^{\circ}$.
A more detailed analysis reveals that the co-polar and cross-polar diagrams shown in Fig.~\ref{fig:df_grd} are rotated by $90^{\circ}$ with respect to each other.

We measured the feedhorn radiation pattern twice, one for each arm of the OMT. The diagram at the left port is the \ang{90} rotation of the diagram at the right port.

\section{Model verification with measurements}
\label{sec:measure}
We verified the results obtained with the electromagnetic analysis of the dual circular polarization system, by performing measurements of the radiation pattern of a system composed by a corrugated feedhorn, a grooved polarizer and a turnstile-junction orthomode transducer.

The angular response characterization has been performed in a microwave anechoic chamber at the Physics Department in the University of Milan. 
The facility we exploited for our measurements is characterized by a working frequency $>\SI{10}{\giga\hertz}$, a dynamics  $>\SI{50}{\decibel}$, an azimuthal range of $\pm\ang{90}$ and automatic positioning and acquisition systems. This facility enables us to make measurements of the principal co-polar and cross-polar planes in the far-field regime. 

\begin{figure}
\centering
\includegraphics[width=0.7\textwidth]{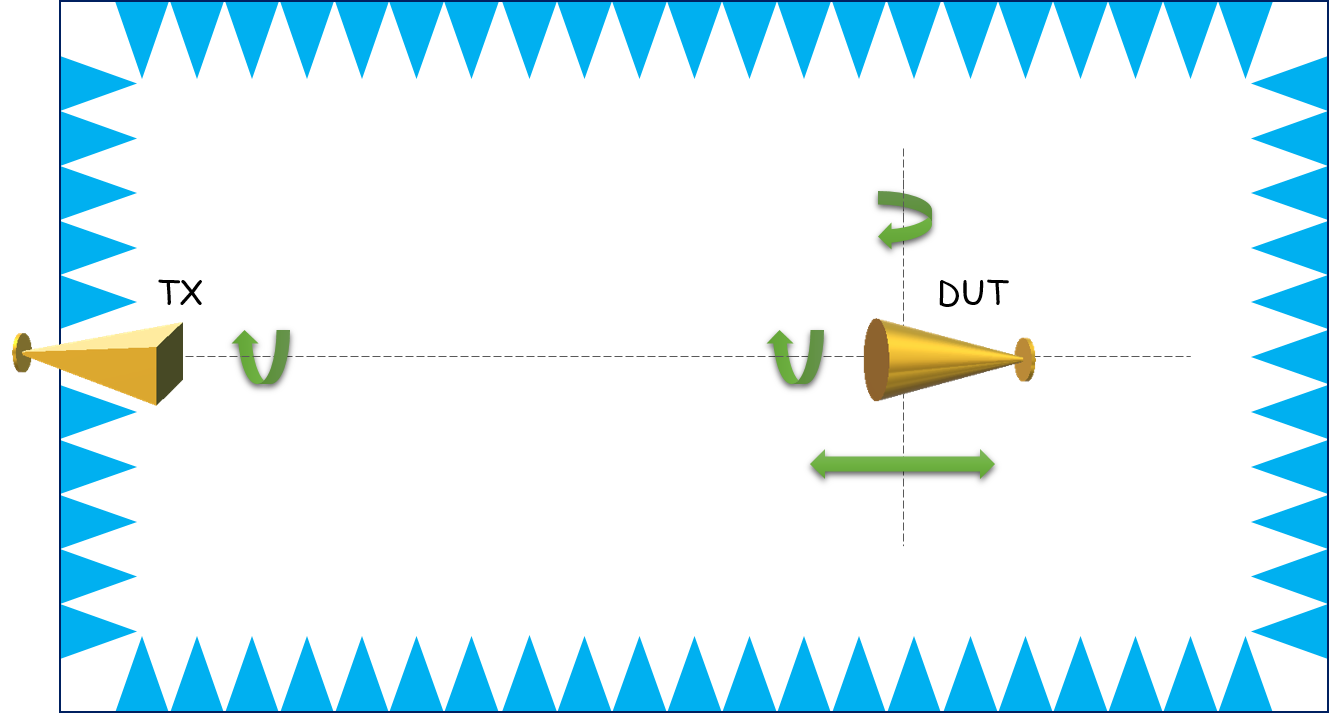}
\caption{Schematics of the anechoic chamber showing the motion axes: DUT polarization, azimuth and center of phase positioning and the TX polarization.}
\label{fig:camera}
\end{figure}

The chamber is provided with five motion axes for the automatic positioning of the transmitter (TX) and the receiver (Device Under Test, DUT) antennas: the rotating axis for the DUT azimuth, the rotating axis for the DUT polarization, the rotating axis for the TX polarization, the linear axis for the DUT center of phase positioning and the linear axes for varying the TX-DUT distance (see Fig~\ref{fig:camera}).

We characterized the radiation pattern of the dual circular-polarization system with several steps, then we compared all the measurements to the simulations to assess their compliance with our expectations.

\begin{figure}
\centering
\includegraphics[width=0.6\textwidth]{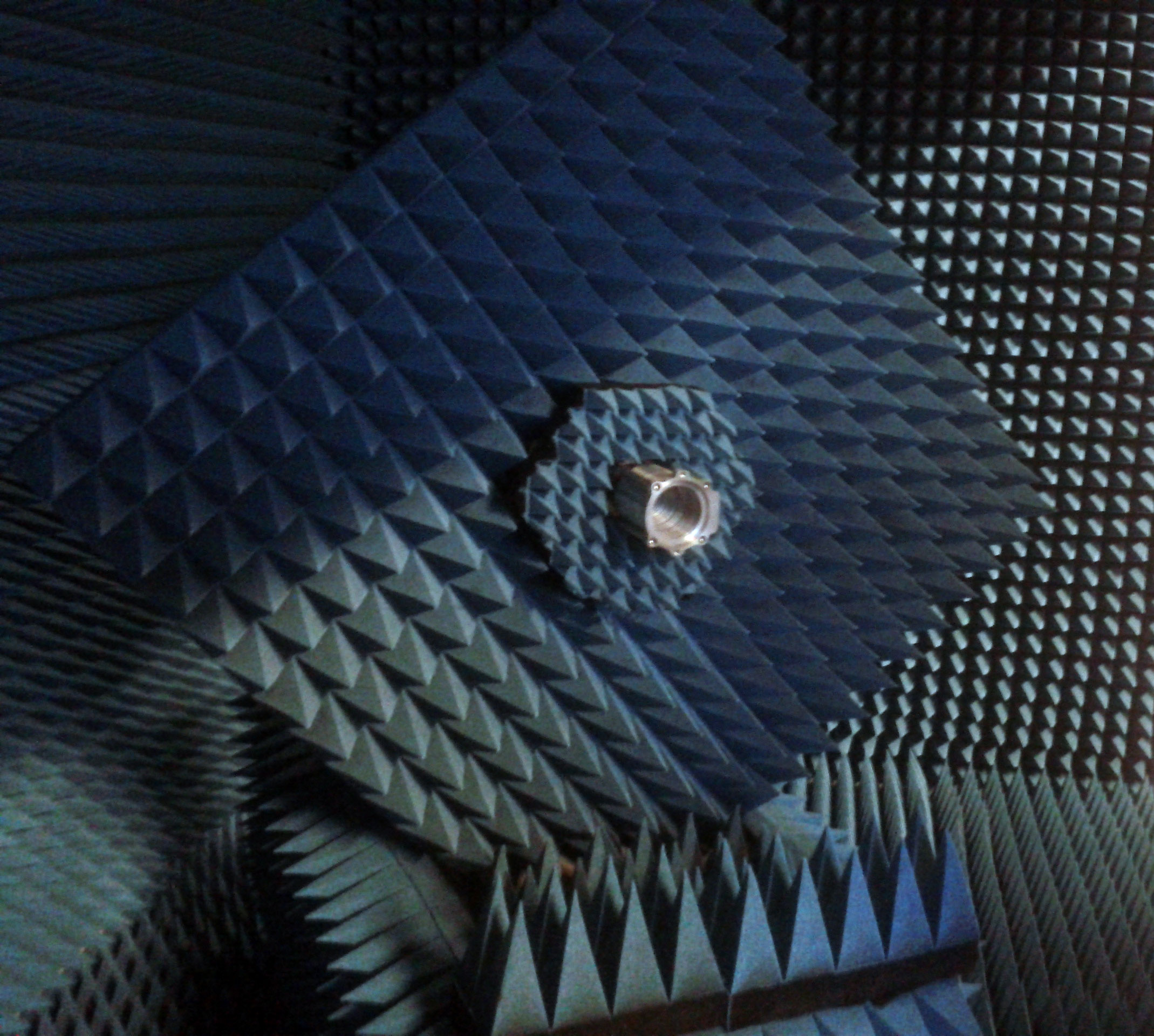}
\caption{Corrugated feedhorn mounted on the DUT mechanical support in the anechoic chamber. The structure is surrounded by Eccosorb shieldings to achieve a cleaner measurement of the radiation patterns.}
\label{fig:setup}
\end{figure}

\begin{figure}
\centering
\includegraphics[width=0.49\textwidth]{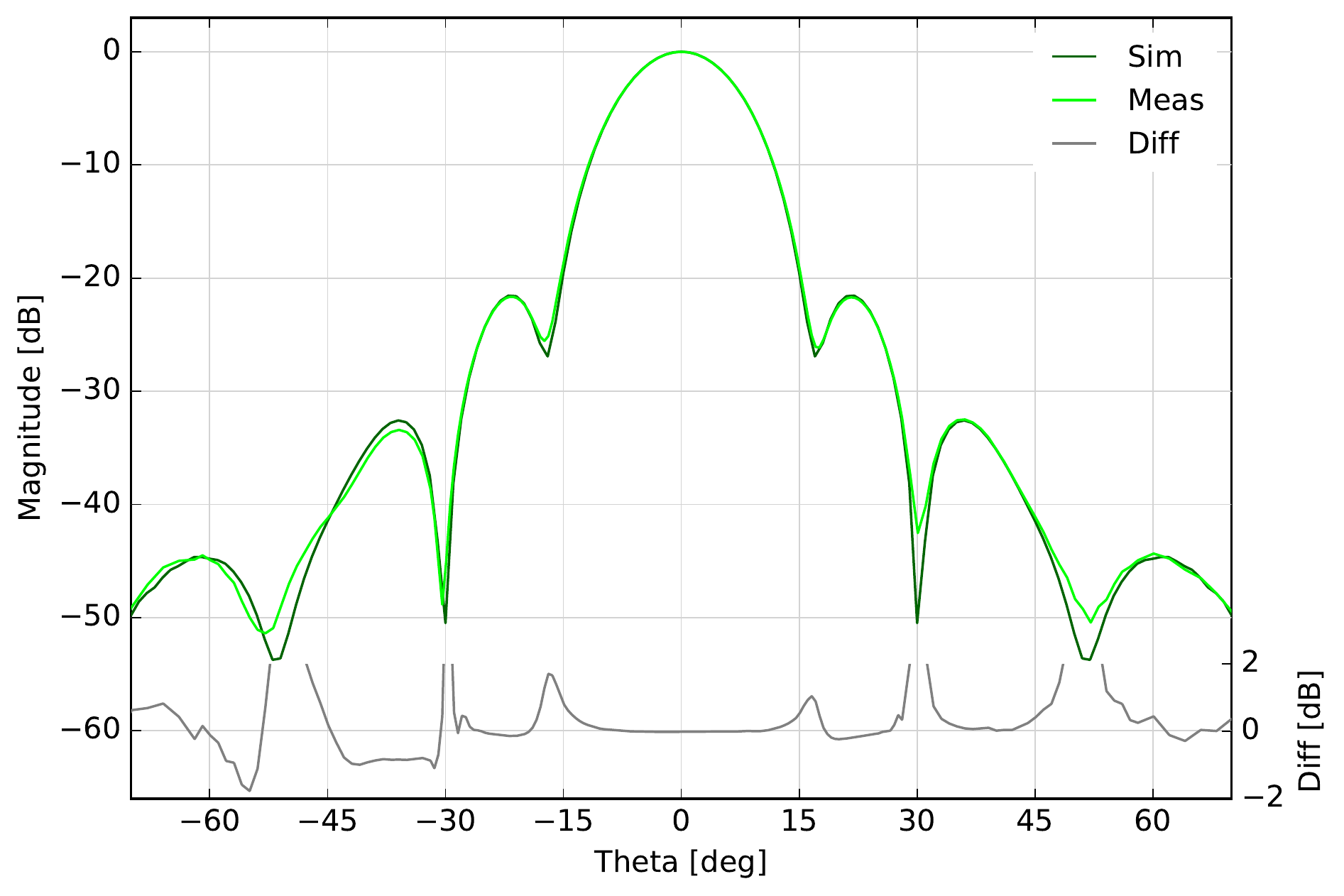}
\includegraphics[width=0.49\textwidth]{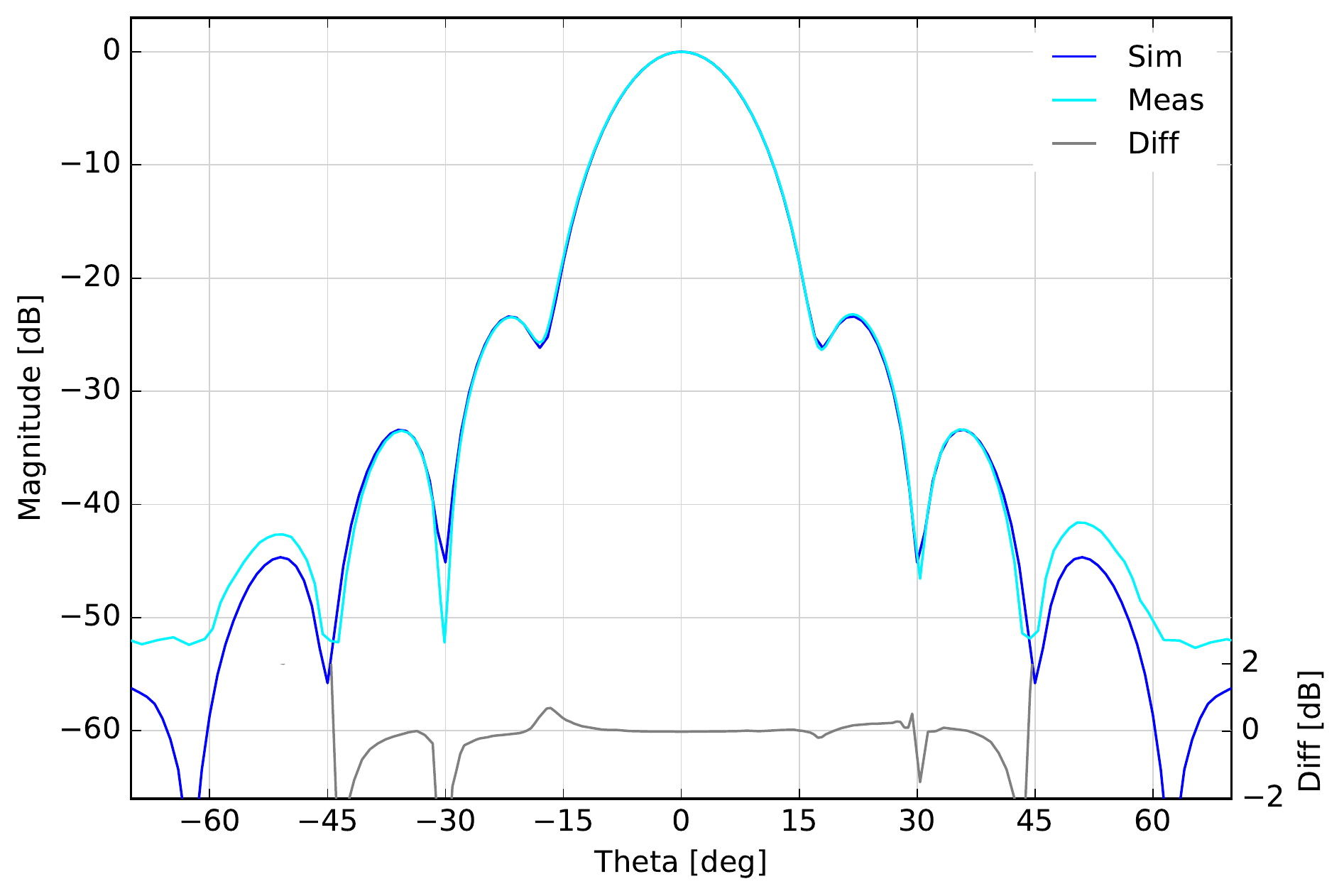}
\includegraphics[width=0.49\textwidth]{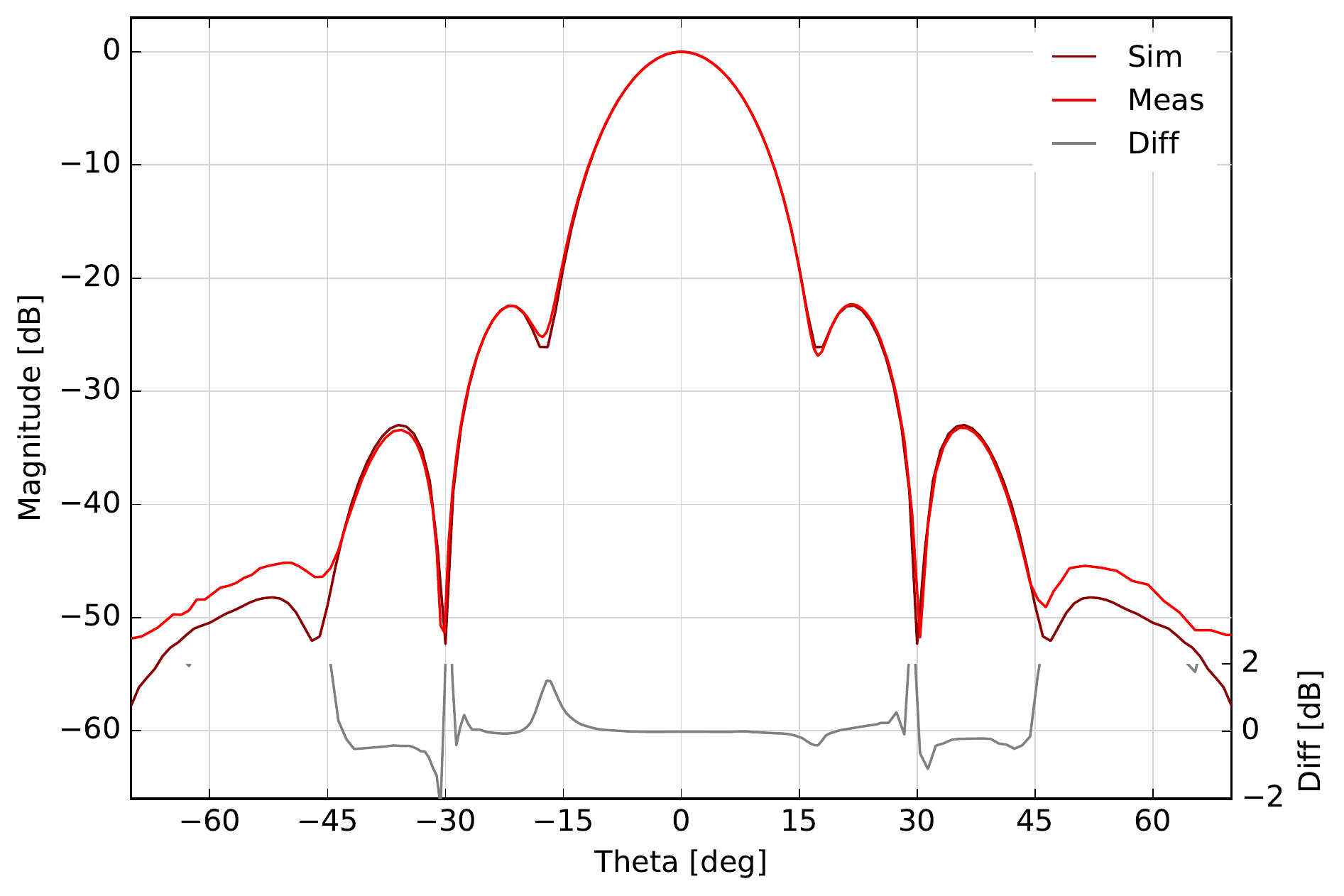}
\includegraphics[width=0.49\textwidth]{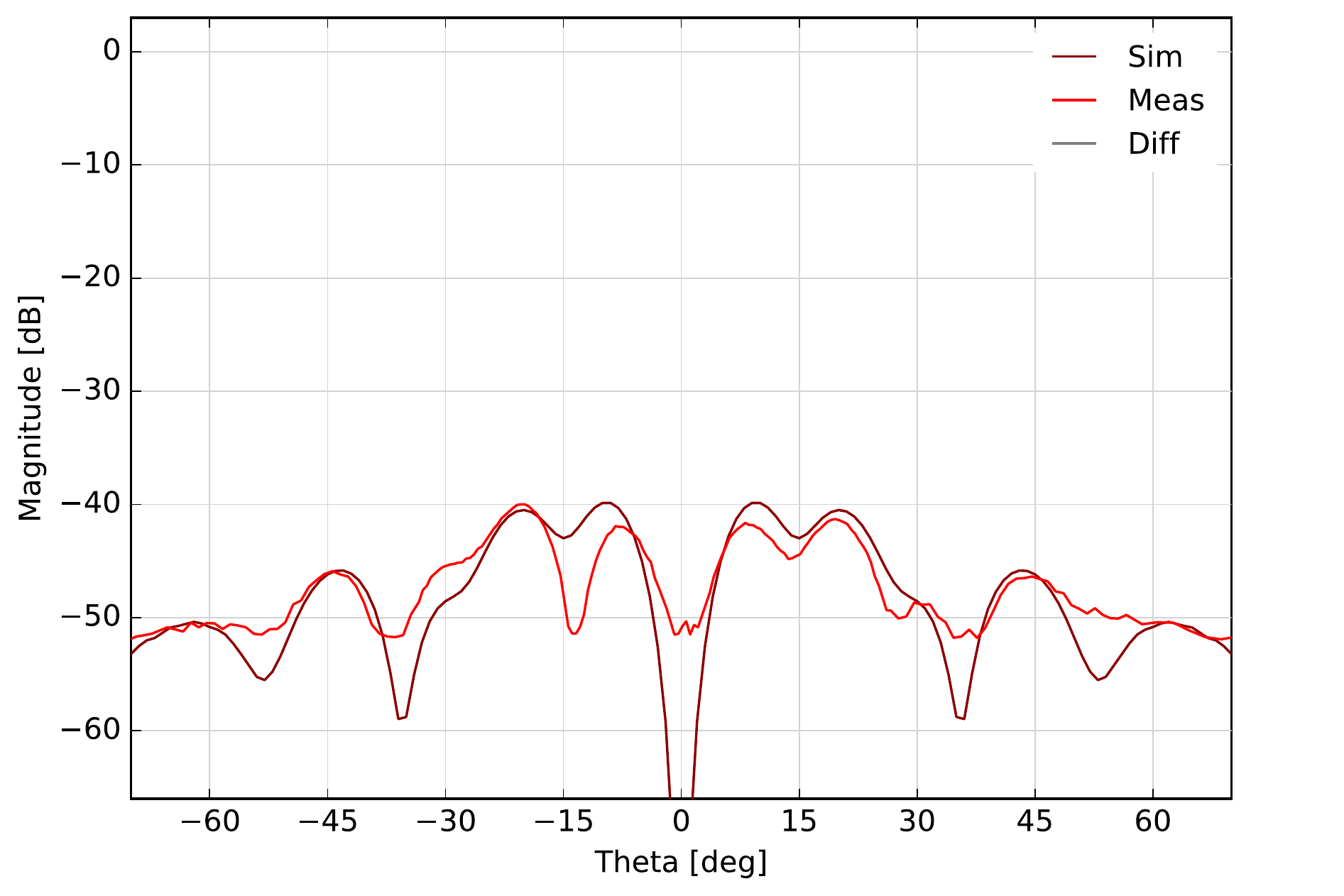}
\caption{Measured radiation patterns at 41.5 GHz compared with the simulation. The difference in magnitude between measured and simulated pattern is reported at the bottom of each plot, with the exception of the cross-polar plane, where systematic effects due to the experimental setup and the lower signal-to-noise ratio determine larger differences compared with the co-polar planes cases. From left to right, \textit{Top}: co-polar E-plane and co-polar H-plane. \textit{Bottom}: co-polar $45^{\circ}$ plane and cross-polar $45^{\circ}$ plane.}
\label{fig:diag_sf}
\end{figure}

We started by measuring the radiation pattern of the feedhorn without the polarizer and OMT.
Fig.~\ref{fig:diag_sf} shows the comparison between measurements and simulations up to the first sidelobe on three co-polar planes and one cross-polar plane: co-polar E-plane, co-polar H-plane, co-polar and cross-polar $45^{\circ}$-plane. All the curves are normalized to unity in a linear scale (0 dB in logarithmic scale). We plot also the difference between the values in dB (grey solid lines) to better understand discrepancies.
Measurements and simulations of the co-polar planes show an agreement within a fraction of dB.
In the whole angular range, we can see that the larger differences appear in the nulls regions, where the signal magnitude rapidly drops by 30 dB or more: in these regions simulations reach about \SI{-50}{\decibel} level, but measurements are limited by the instrumental noise.
All antenna radiation patterns have been measured at the frequency 41.5 GHz, which is the central frequency of the feedhorn.

\begin{figure}[htp]
\centering
\includegraphics[width=0.49\textwidth]{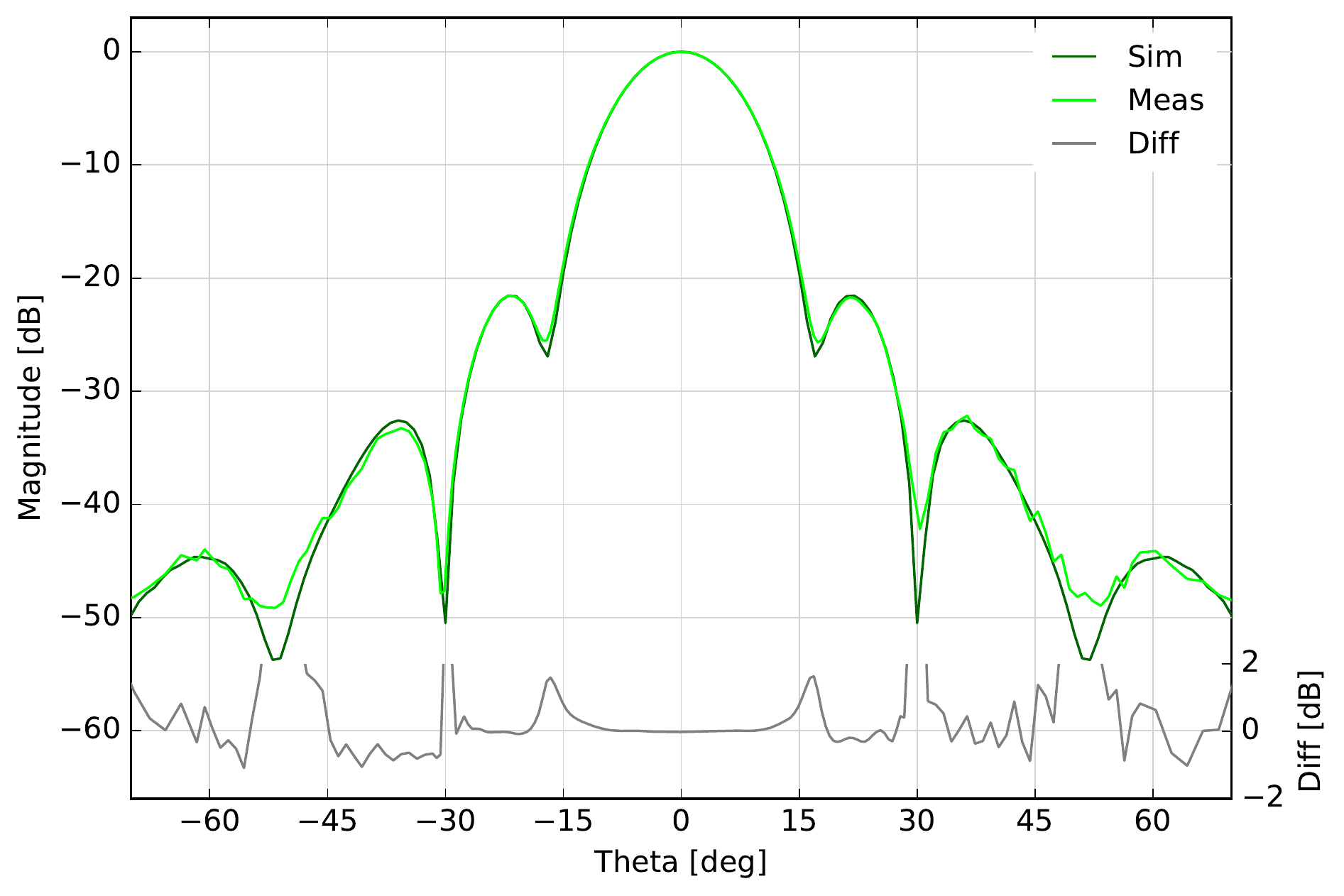}
\includegraphics[width=0.49\textwidth]{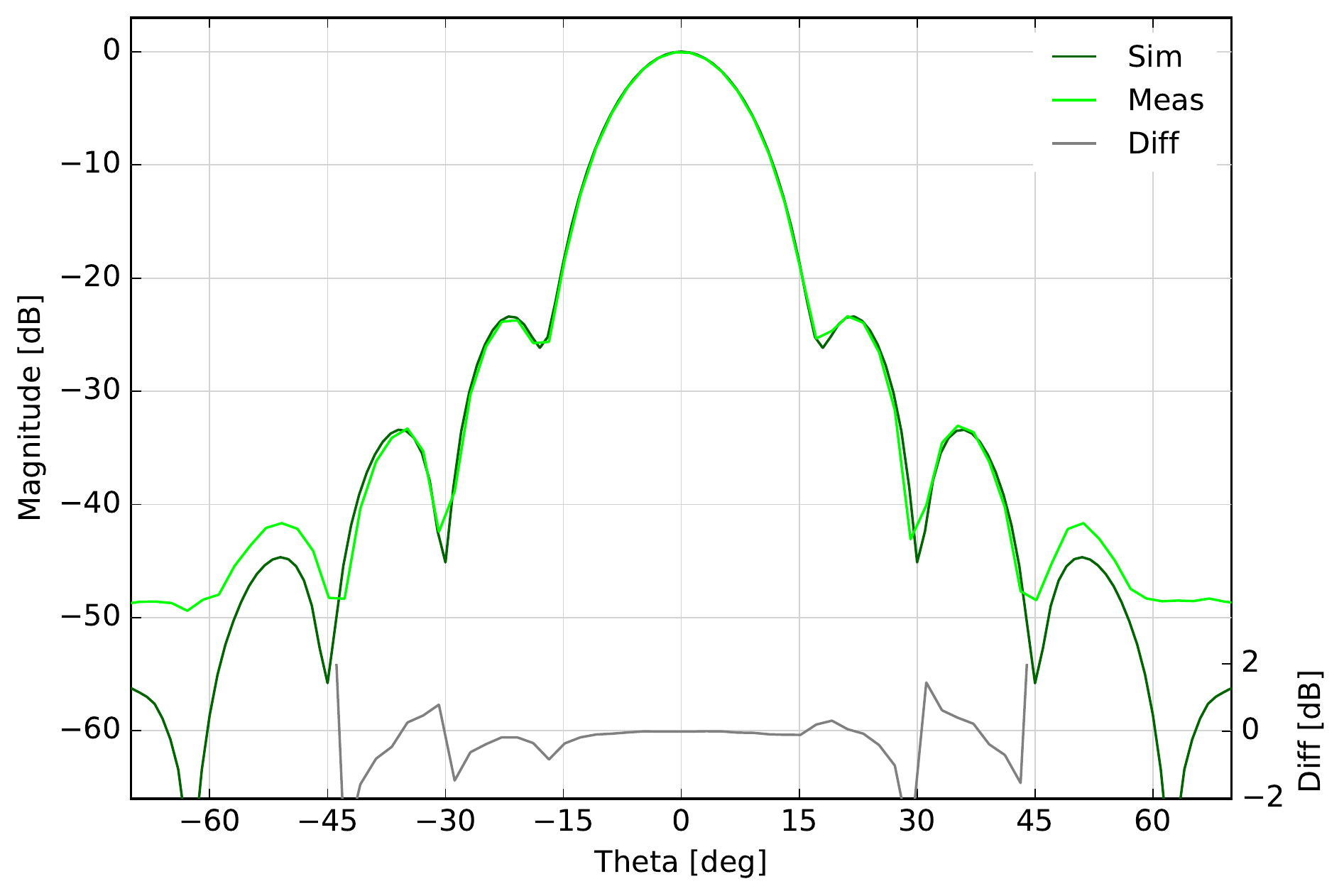}
\includegraphics[width=0.49\textwidth]{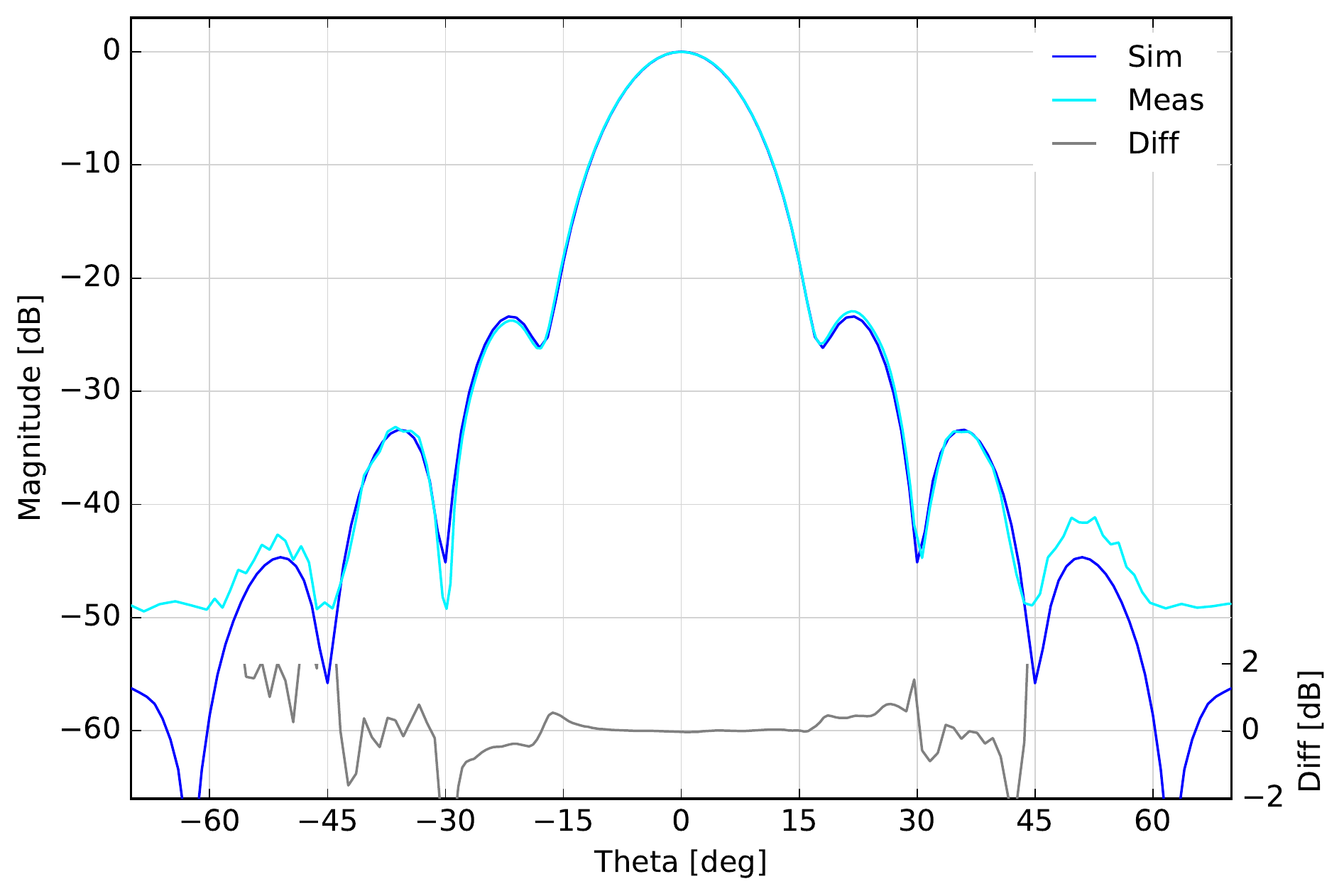}
\includegraphics[width=0.49\textwidth]{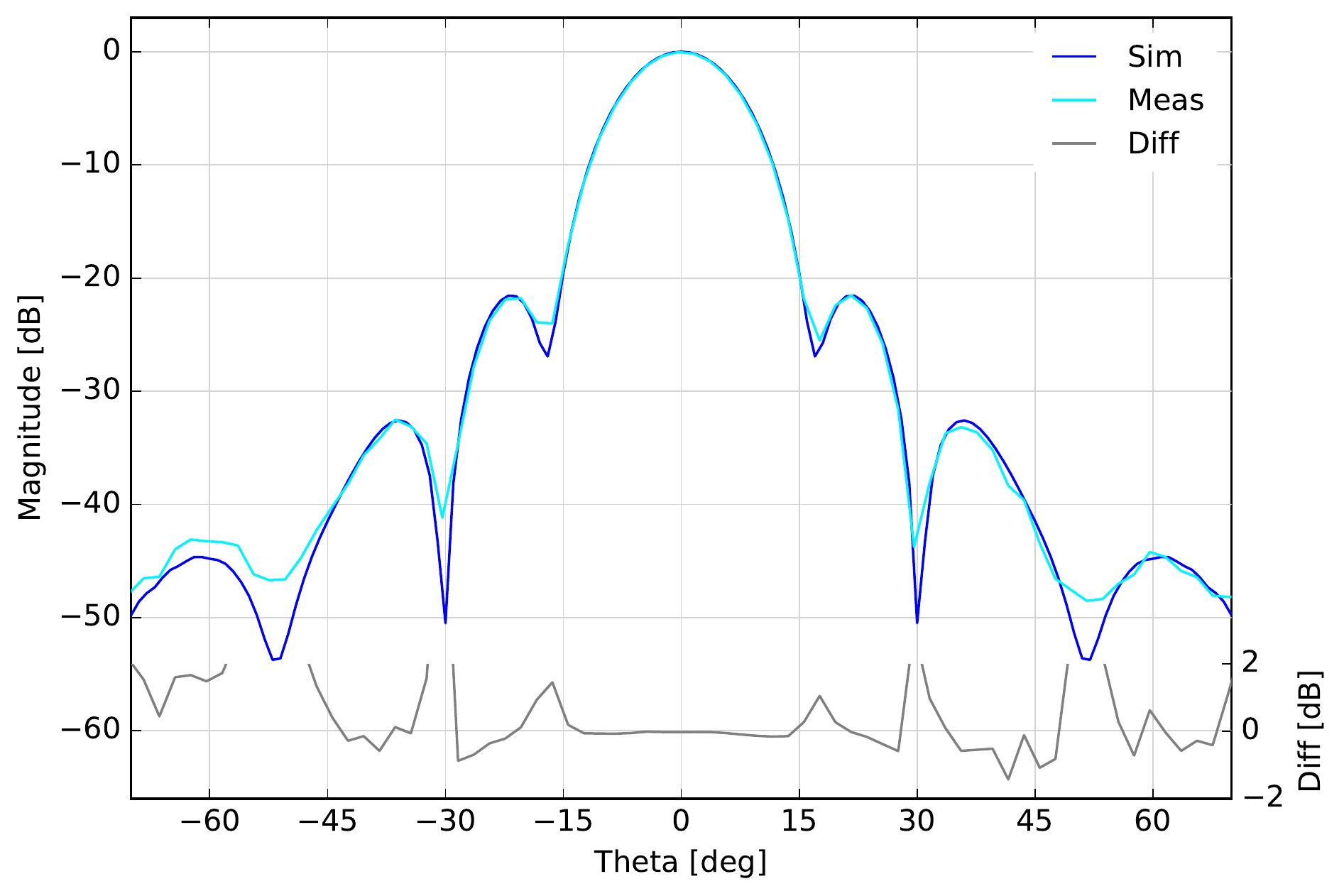}
\includegraphics[width=0.49\textwidth]{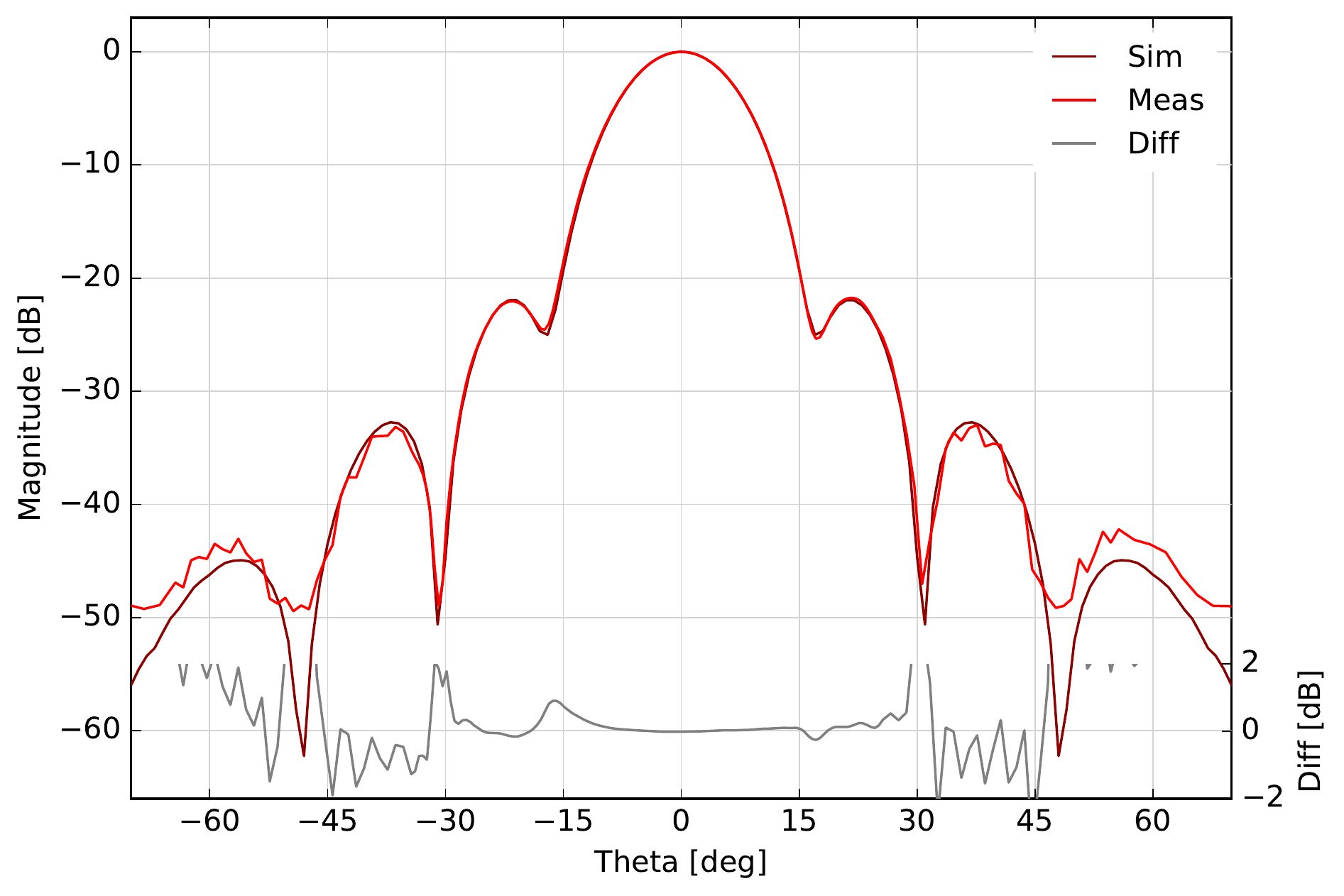}
\includegraphics[width=0.49\textwidth]{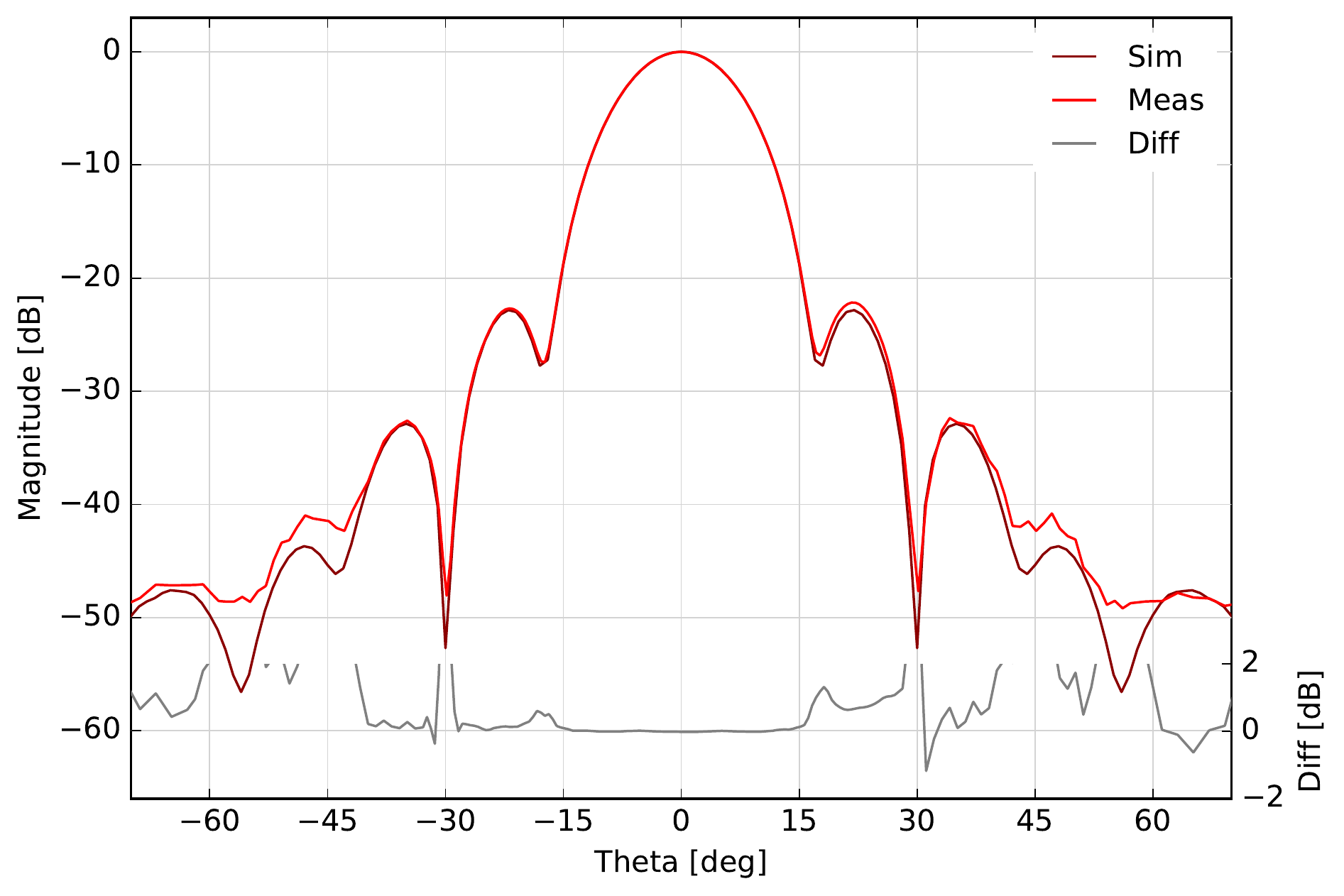}
\caption{Circular waveguide system (feedhorn, polarizer, OMT) measured radiation patterns at 41.5 GHz compared with the simulations. The difference in magnitude between measured and simulated pattern is reported at the bottom of each plot. From left to right, \textit{Top}: co-polar E-plane and cross-polar E-plane. \textit{Middle}: co-polar H-plane and cross-polar H-plane. \textit{Bottom}: cross-polar \SI{45}{\degree} plane and cross-polar \SI{45}{\degree} plane.}
\label{fig:meas_pol_omt}
\end{figure}

Then, we proceeded with the measurement of the whole circular waveguide system, adding the polarizer and the OMT to the RF chain. Since the OMT has two ports, we picked the signal from one port, while matching the other one with a load.
Fig.~\ref{fig:meas_pol_omt} shows the comparison between measurements at one port of the OMT and simulations on three co-polar planes and three cross-polar planes: co-polar E-plane, cross-polar E-plane, co-polar H-plane, cross-polar H-plane, co-polar $45^{\circ}$-plane and cross-polar $45^{\circ}$-plane.

Measurements and simulations show an agreement within a fraction of dB in the angular region up to the second lobe after the main beam. We report the difference in magnitude between the measured and simulated pattern at the bottom of each plot. As we expected, the cross-polar patterns show a co-polar shape, with the same peak value, and the co-polar E-plane and the cross-polar H-plane overlap. The same applies to the co-polar H-plane and the cross-polar E-plane. 

\begin{figure}[htp]
\centering
\includegraphics[width=0.48\textwidth]{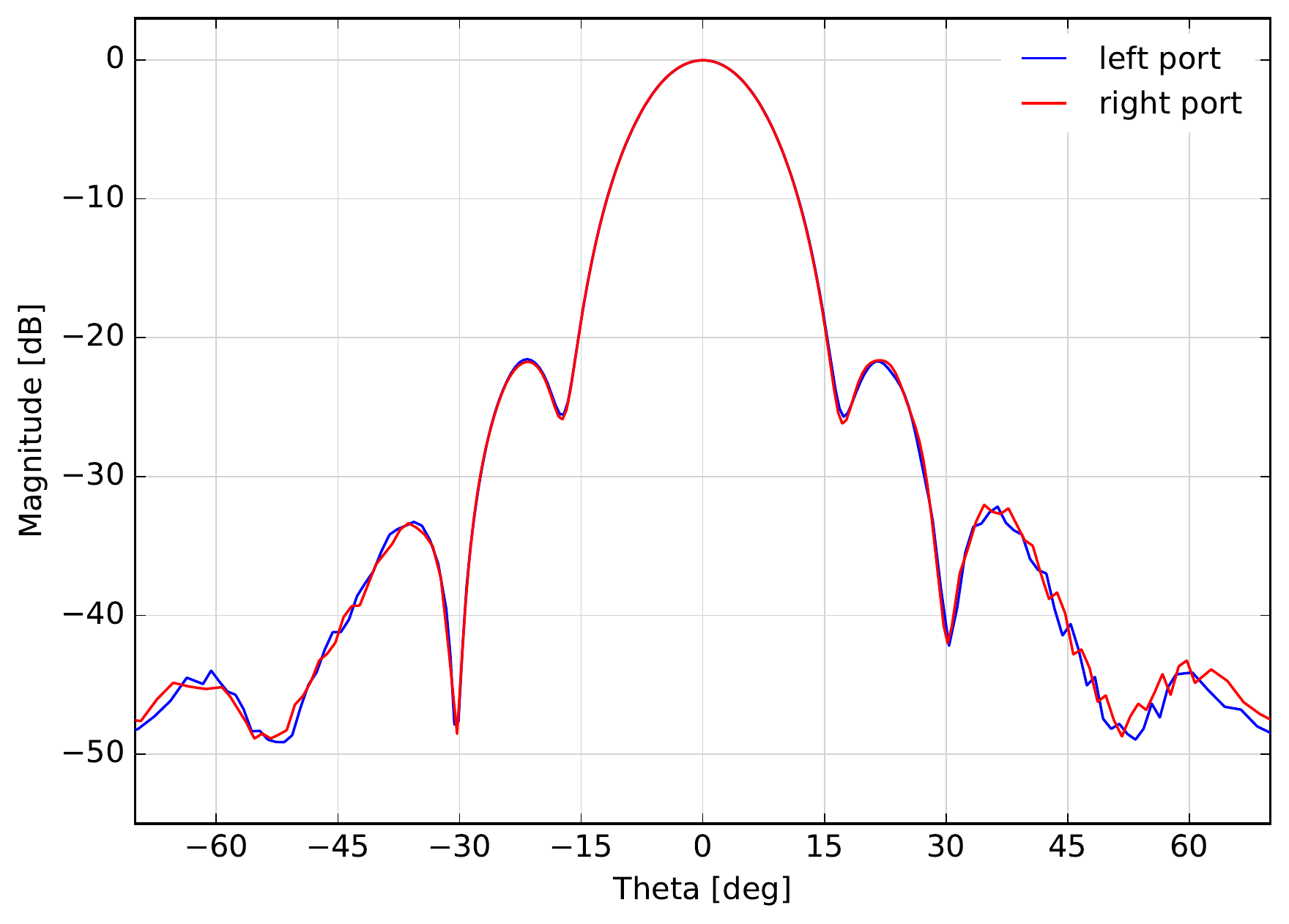}
\includegraphics[width=0.48\textwidth]{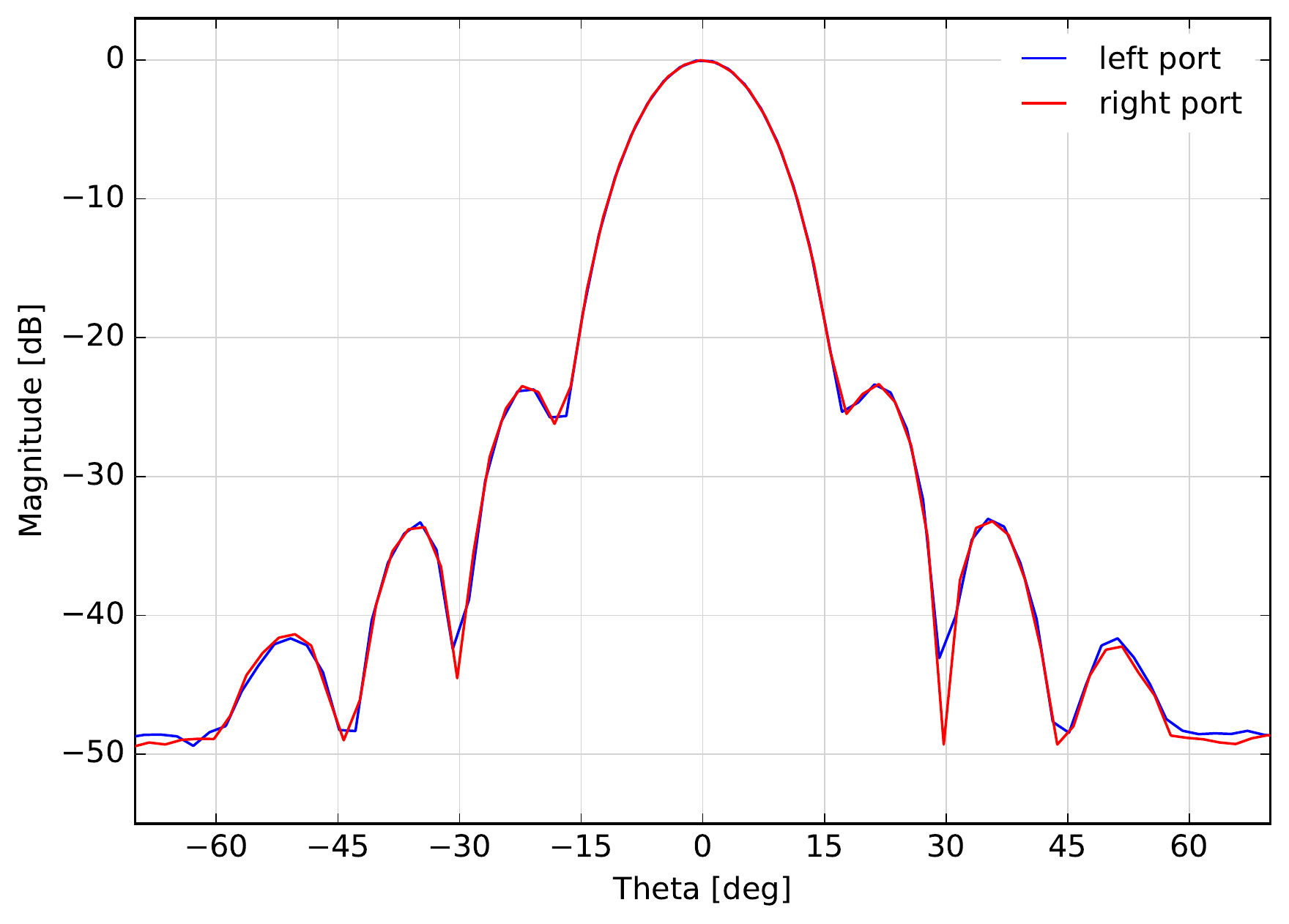}
\includegraphics[width=0.48\textwidth]{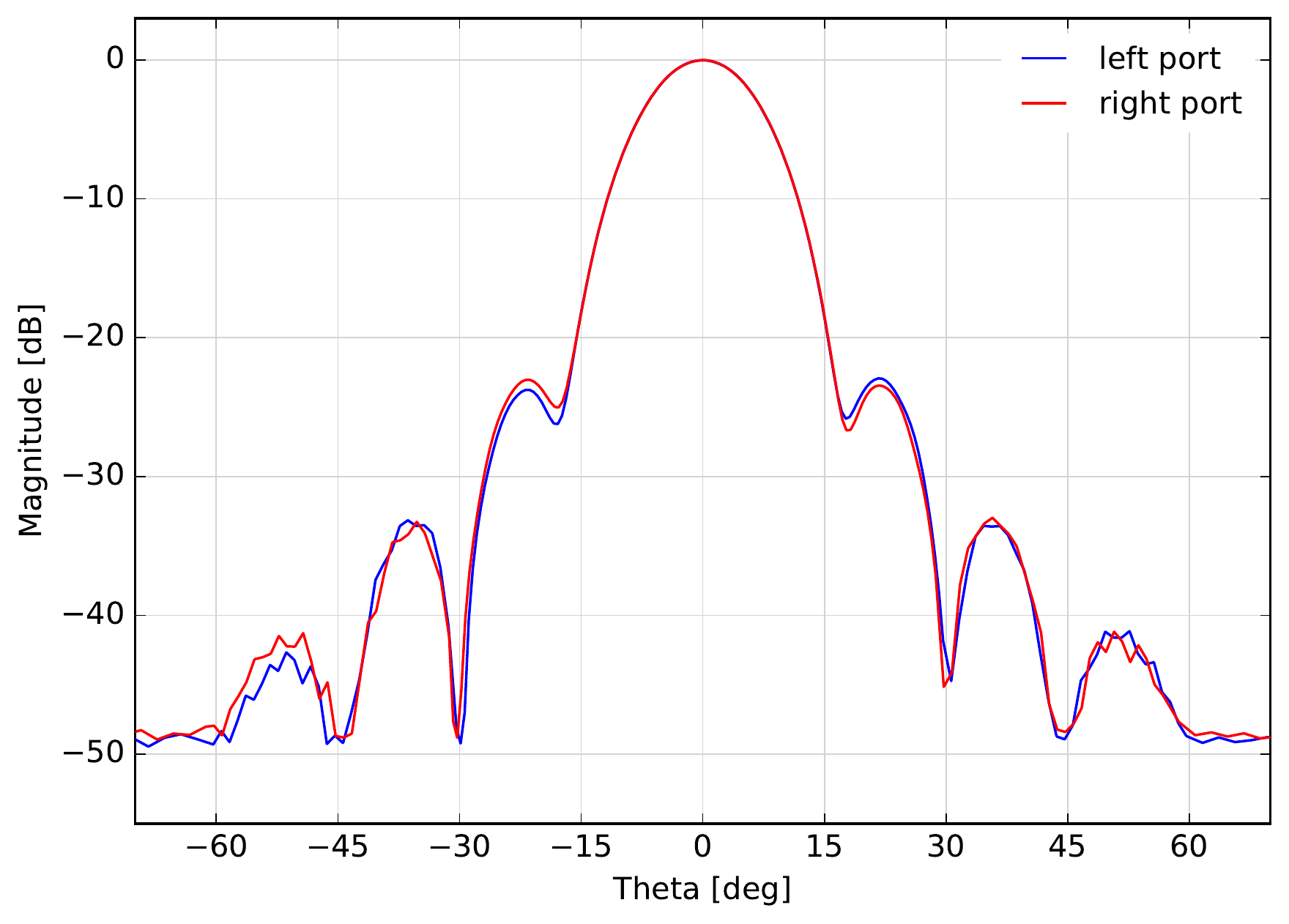}
\includegraphics[width=0.48\textwidth]{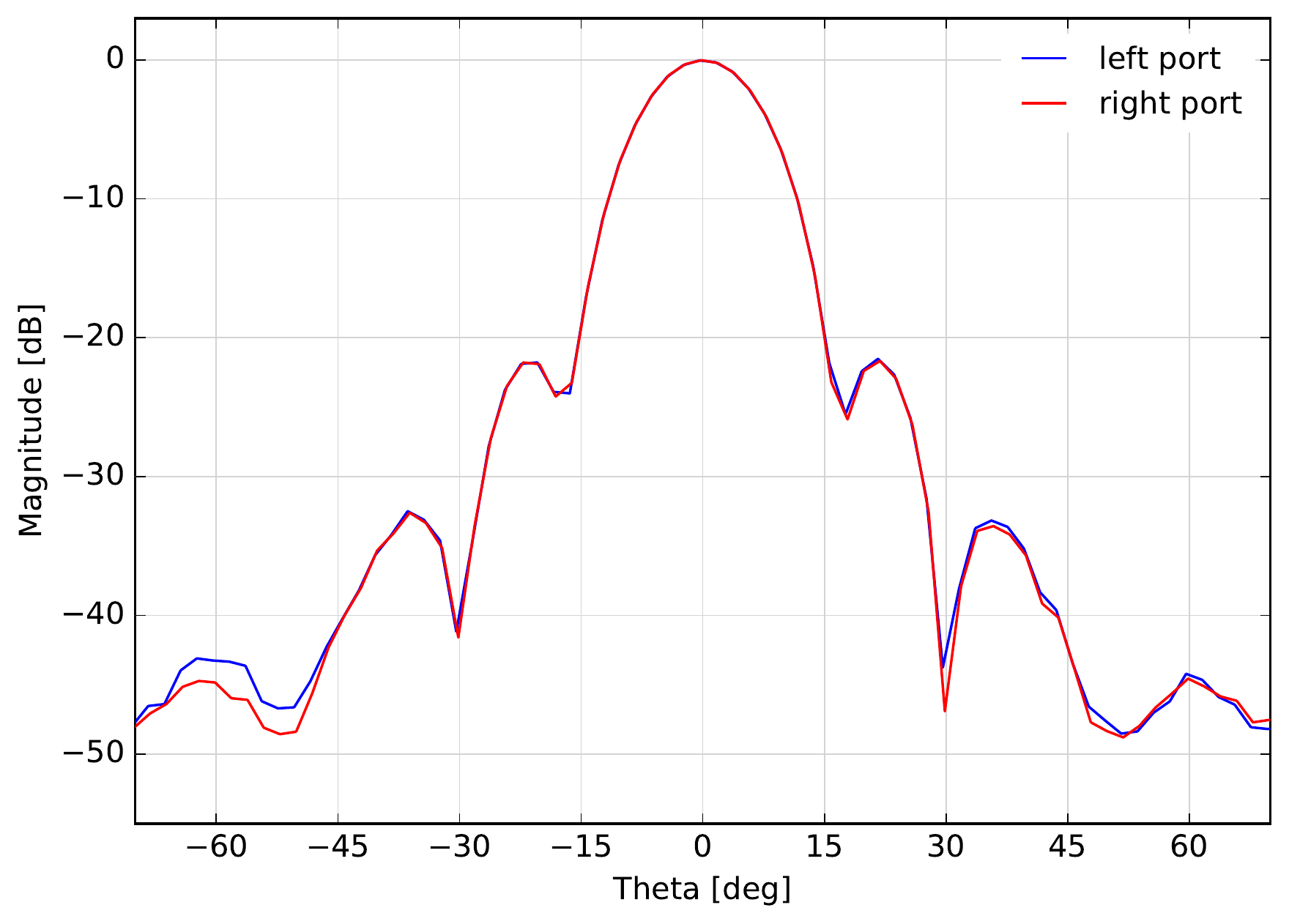}
\includegraphics[width=0.48\textwidth]{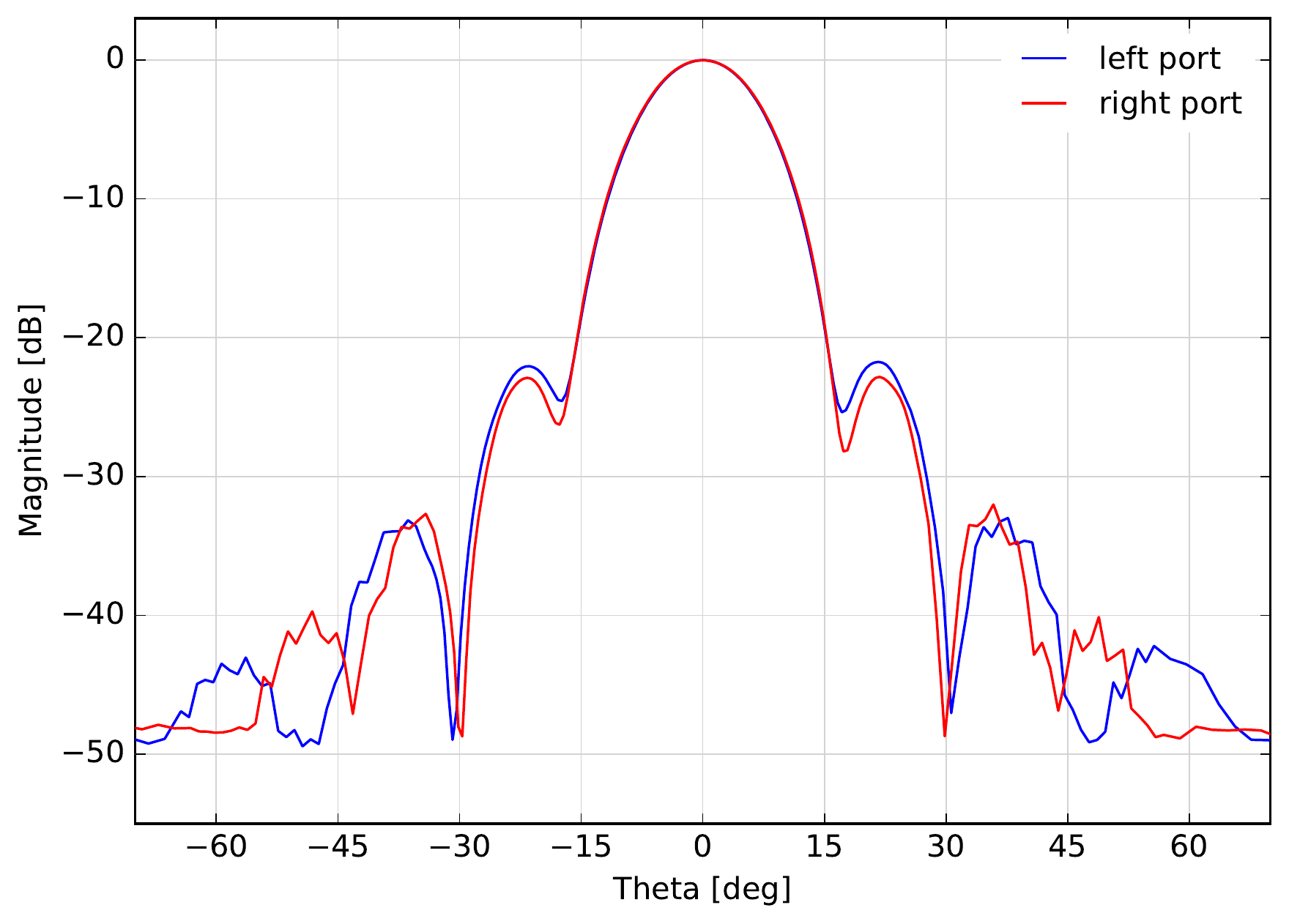}
\includegraphics[width=0.48\textwidth]{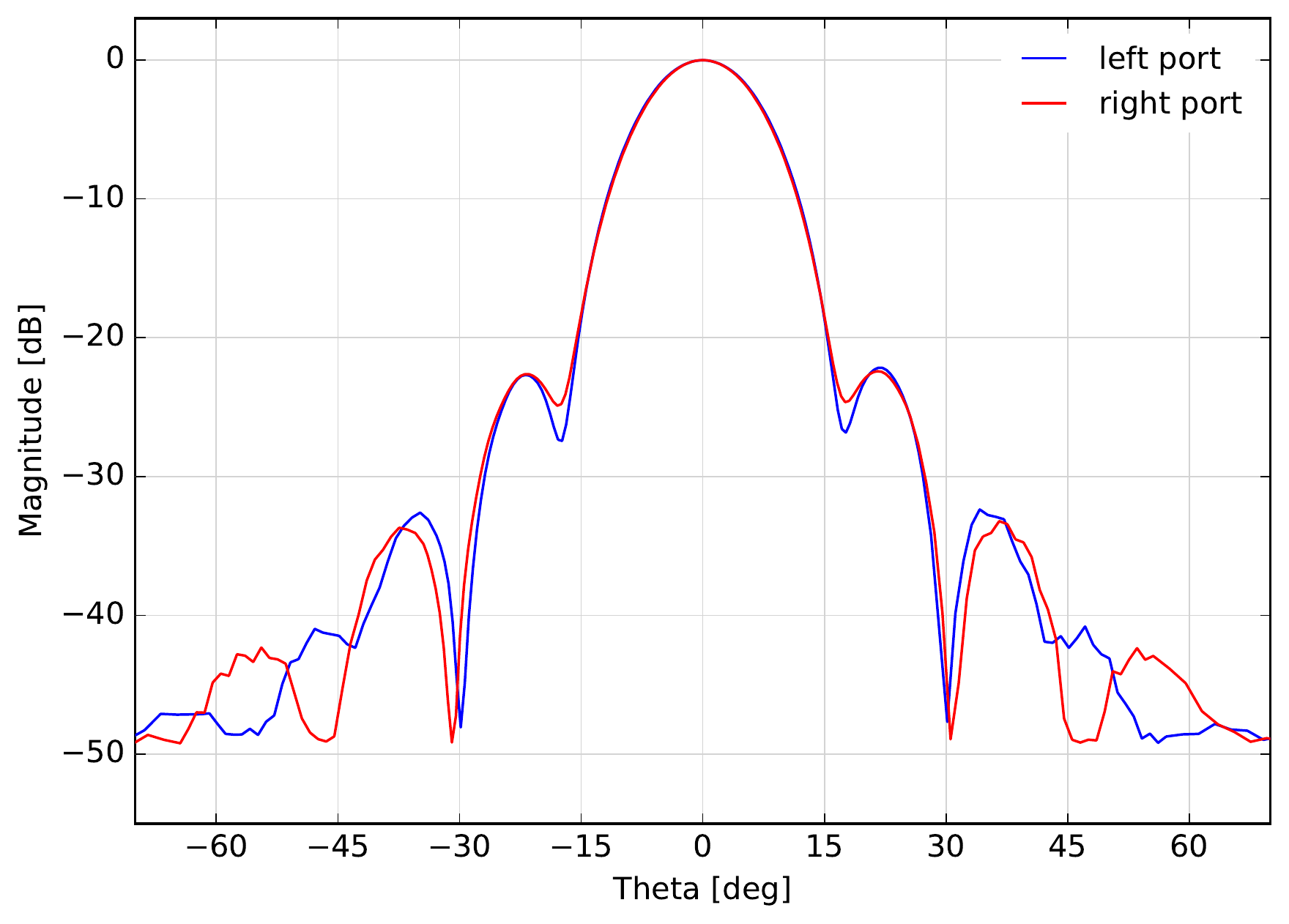}
\caption{Comparison of the measured radiation patterns at the two OMT ports. From left to right, \textit{Top}: co-polar E-plane and cross-polar E-plane. \textit{Middle}: co-polar H-plane and cross-polar H-plane. \textit{Bottom}: cross-polar \SI{45}{\degree} plane and cross-polar \SI{45}{\degree} plane. As expected, the diagram at the left port is the \SI{90}{\degree} rotation of the right port one.}
\label{fig:porte}
\end{figure}

Finally, we repeated the measurement for both OMT ports. The comparison between the planes measured at the two ports is shown in Fig.~\ref{fig:porte}. Let us focus on the diagrams relative to the 45$^{\circ}$ plane (third row in Fig.~\ref{fig:porte}. We can see that the red and blue curves are swapped, which means that the co-polar component at one port becomes the cross-polar component at the other port (and vice versa), as we expected since the diagram at one of the ports is the \SI{90}{\degree} rotation of the diagram at the other port, as shown also in Fig.~\ref{fig:df_grd}.

The agreement between measurements and simulations verifies the model described in paragraph \ref{sec:model}. 
Furthermore, it tells us that the optics analysis made considering the feedhorn as source of a linearly polarized signal is efficient.  

\section{Conclusions}
In this paper, we presented the electromagnetic model of a dual circular polarization antenna-feed system, consisting of a corrugated feedhorn, a polarizer and an orthomode transducer, that is an architecture frequently used in microwave telescopes. For example, experiments for CMB polarization observation, based on coherent receivers, largely employ dual circular polarization passive front-ends, so that this work provides a useful description and modelling of such systems. The model has been implemented with Ticra GRASP software by superposing two linearly polarized feedhorns with a $\pi$/2 phase difference. In this way we can take into account the effect of the polarizer, that behaves differently for the two polarizations of the incoming electric field.

The model has been verified by means of radiation pattern measurement, performed in the anechoic chamber at the Physics Department of the University of Milan. The measurement has been performed for both arms of the orthomode transducer, to check that the diagram at one port is the 90$^{\circ}$ rotation of the diagram at the other port. Simulations and measurements show an agreement at the level of fraction of a dB up to the first sidelobes, thus verifying the model.

\acknowledgments
Authors would like to acknowledge Renzo Nesti (Osservatorio Astrofisico di Arcetri, Firenze), Oscar Antonio Peverini (IEIIT, Torino) and Francesco Del Torto (formerly, Universit\`a degli Studi di Milano) for long, as well as precious, discussions on the matter of this paper.

\bibliographystyle{JHEP}

\end{document}